\documentclass[a4paper,10pt]{article}
\usepackage{graphicx}

\begin{document}

\title{The young open cluster Trumpler 3}
\author{G. Maciejewski$^{1,2}$ and {\L}. Bukowiecki$^1$\\
\small $^1$Toru\'n Centre for Astronomy, N. Copernicus University, Gagarina 11, PL-87100 Toru\'n, Poland\\ 
\small $^2$Astrophysikalisches Institut und Universit\"ats-Sternwarte, Schillerg\"asschen 2-3, D-07745 Jena, Germany\\
\small e-mail: gm@astri.uni.torun.pl
}
\maketitle

\textit{Abstract}: We present a photometric and spectroscopic study of the poorly investigated open cluster Trumpler~3. Basic parameters such as the age of $70\pm10$ Myr, the color excess $E(B-V)=0.30\pm0.02$ mag, the distance of $0.69\pm0.03$ kpc and the limiting radius of 12' were redetermined and compared with previous preliminary studies. The distance of $0.65\pm0.09$ kpc was determined independently by spectral parallaxes. Simultaneously, our analysis allowed us to estimate a total number of members to be $N_{\rm{tot}}=570\pm90$ and a total mass of the cluster to be $M_{\rm{tot}}=270\pm40$ $\rm{M}_{\odot}$. We also determined a state of cluster's dynamical evolution. We conclude that Trumpler~3 is a young low-massive stellar ensemble with a typical mass function slope, located near to the outer edge of the Galaxy's Orion Spur. As a result of a wide-field search for short period variable stars, 24 variables were discovered in the cluster's area. Only one of them -- a variable of the $\gamma$-Dor type -- was found to be a likely cluster member.
 

\section{Introduction}

\begin{figure}
 \includegraphics[width=8.3cm]{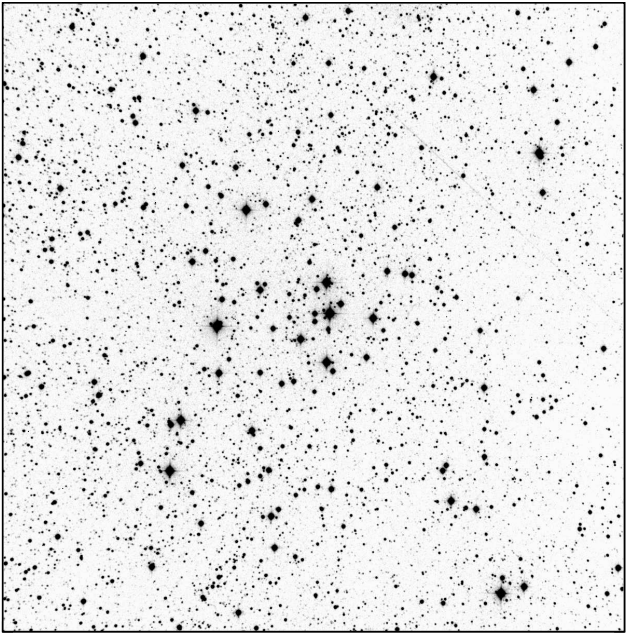}
 \caption{A $ 43'$ $\times$ $43'$ field around the open cluster Trumpler~3. North is up, East to the left.}
 \label{fig1}
\end{figure}

Among Galactic open clusters there are objects which basic parameters such as an age or distance still remain unknown. In this paper we present photometric and spectroscopic investigations of Trumpler~3 (C~0307+630 = Harvard~1) -- a poor, low-contrast open cluster located in the constellation of Cassiopeia (Fig.~\ref{fig1}). 

The existence of an open cluster in the area occupied by Trumpler~3 was postulated by Barnard at al. (1927). Trumpler (1930) described the cluster as a very loose open cluster, not rich but regular in outline and structure. Moreover, the author classified this stellar ensemble as a cluster of type II3p and determined the angular diameter of 17' and the distance of 0.69 kpc. Ruprech (1966) classified Trumpler~3 as a poor cluster of class III3p. According to the \textit{New catalogue of optically visible open clusters and candidates} by Dias et al. (2002), the angular diameter of the cluster is 14'. Kharchenko at al. (2005) performed the first photometric studies of the cluster but they were limited to stars brighter than $V=12-13$ mag. The authors obtained the limiting radius of 30', the core radius of $\sim$8', the age of $\sim$220 Myr, and the apparent distance modulus of 8.95 mag which they converted to the distance of 450 pc. Moreover, the interstellar reddening $E(B-V)$ was found to be 0.22 mag. Mermilliod at al. (2008) measured the radial velocities for two stars -- members of the cluster -- and determined the mean radial velocity of the cluster $V_{\rm{r}}=-8.58\pm1.12$ km s$^{-1}$. 

No dedicated CCD studies of Trumpler~3 have been performed to date. To fulfill this niche, we gathered wide-field deep $BV$ photometry and combined it with near-infrared $JHK_{\mathrm{S}}$ photometry extracted from the 2-Micron All Sky Survey (2MASS, Strutskie et al.~2006). Our analysis resulted in determination of structural and fundamental astrophysical parameters, as well as in estimating a total number of members and total mass of the cluster. The low-resolution prism spectra were gathered for brightest stars what allowed us to determine independently the distance to the cluster by spectral parallaxes. Moreover, the cluster's field was searched for variable stars -- possible cluster members.

\section{Observations and data reduction}

Observations were performed with the 90/180 cm Schmidt-Cassegrain Telescope (TSC90) located at the Astronomical Observatory of the Nicolaus Copernicus University in Piwnice near Toru\'n, Poland. The telescope was equipped with SBIG STL-11000 CCD camera with KAI-11000M CCD detector (4008 $\times$ 2672 pixels $\times$ 9 $\mu$m). The field of view of the instrument was 72' in declination and 48' in right ascension with the scale of 1.08 arcsec per pixel. 

The photometric frames were obtained in a direct imaging mode. A $2 \times 2$ binning was applied to increase the signal-to-noise ratio. That resulted in the real scale of 2.15 arcsec per pixel. The spectroscopic frames were obtained in a prism mode with a 60-cm BK7-glass objective prism. In this case no binning was used. The CCD frames were processed using a standard procedure that included subtraction of dark frames and flat-fielding with twilight flats.

\subsection{Deep $BV$-band photometry}

The $BV$-band observations were gathered on October 6 and 30, 2005. During the first run 2 short, 2 medium, and 2 long exposures were acquired in each filter with exposure time of 10 s, 60 s, and 600 s, respectively. In second night only a set of short and long observations were acquired. Additionally, 5--6 observations of Landolt's (1992) calibration field were carried out at various airmasses to monitor atmospheric condition and to determine atmospheric extinction coefficients. Only night on October 30 was found to be photometric and data from this night were used to obtain absolute photometry of the cluster field. In next step these data were used to calibrate photometry gathered on October 6.  

The atmospheric extinction coefficients in a given filter, $k_{\rm{V}}$ and $k_{\rm{B}}$, were determined using 5 observations of the Landolt's (1992) calibration field SA 95 at airmasses $X$ between 1.6 and 2.7. Typically, more than 150 stars were detected in every frame. Their instrumental magnitudes were used to determine $k$. A median value was taken as the one best representing a night. We got $k_{\rm{V}}=0.294$ and $k_{\rm{B}}=0.412$.

The raw instrumental magnitudes $b_{\rm{raw}}$ and $v_{\rm{raw}}$ of stars in the Landolt field were derived with the aperture photometry with the software pipeline developed for the Semi-Automatic Variability Search sky survey (SAVS, Niedzielski et al. 2003). Instrumental magnitudes outside the atmosphere $b$ and $v$ were calculated as
\begin{equation}
 b=b_{\rm{raw}}-k_{\rm{B}}X\, , \;
\end{equation}
\begin{equation}
     v=v_{\rm{raw}}-k_{\rm{V}}X\, . \;
\end{equation}
The calibration coefficients that transform instrumental magnitudes into standard ones were determined using 36 standard stars detected in the calibration field and covering the $(B-V)$ color index in the range between -0.215 and 1.999 mag. The following equations were derived
\begin{equation}
 V-v=(-0.10\pm0.02)(b-v)+(20.297\pm0.015)\, , \;
\end{equation}
\begin{equation}
     B-V=(1.17\pm0.03)(b-v)+(0.14\pm0.01)\, , \; 
\end{equation}
where $B$ and $V$ are standard magnitudes and $b$ and $v$ are mean instrumental magnitudes corrected for the atmospheric extinction. The median values of residuals for $V$ and $(B-V)$ are 0.017 mag and 0.014 mag, respectively. 

The magnitudes in the cluster field were obtained by point-spread function (PSF) fitting. We used DAOPHOT package (Stetson 1987) implemented in IRAF\footnote{IRAF is distributed by the National Optical Astronomy Observatories, which are operated by the Association of Universities for Research in Astronomy, Inc., under cooperative agreement with the National Science Foundation.}. The cluster frames were limited to $43' \times 43'$ around the cluster center (Fig.~\ref{fig1}) due to deformations of stellar profiles in outer parts of the field of view. To construct the PSF empirically, we selected 16-23 bright isolated stars. A second-order variable PSF was used to take the systematic pattern of PSF variability with position on the chip into account. Aperture corrections were determined using aperture photometry of stars which were used while building a PSF profile in every frame. The instrumental coordinates of stars in frames were transformed into equatorial ones making use of positions of stars brighter than 17.5 mag and extracted from the Guide Star Catalog.

The final list of stars contains equatorial coordinates, $V$ magnitudes, and $(B-V)$ color indices. It is available in electronic form at the survey web site\footnote{http://www.astri.uni.torun.pl/\~{}gm/OCS} and the WEBDA\footnote{http://www.univie.ac.at/webda/} database (Mermilliod 1996).

\subsection{Prism spectroscopy}

Spectroscopic data were gathered on February 18, 2007. To cover the whole cluster field, six partially overlapping fields were observed. For each one a set of one short (240 s) and one long (960 s) exposures was secured. Individual spectra were extracted with IRIS\footnote{http://www.astrosurf.com/buil} and then processed with Visual Spec\footnote{http://www.astrosurf.com/vdesnoux}. The spectra cover a wavelength range of 3900--8300~\AA~ with a resolution of about $\sim$5~\AA~ and $\sim$20~\AA~ at H$_{\gamma}$ and H$_{\alpha}$, respectively.

\subsection{Wide-field monitoring}

The field of Trumpler~3 was search for variable stars during 15 nights spread between October 6, 2005 and April 29, 2009. As a result of 53 hours of monitoring, we collected 91 and 182 long-exposure (400-600 s) frames in the $B$ and $V$ bands, respectively. To investigate two dozens of the brightest stars, we gathered simultaneously 43 and 379 short-exposure (20 s) frames in $B$ and $V$ bands, respectively. The whole available field of view of the telescope (i.e. $48' \times 72'$) was used because magnitudes were derived with the differential aperture photometry. Five comparison stars with magnitudes $13.6<V<14.0$ and four with magnitudes $11.5<V<12.0$ were chosen for long and short exposures, respectively. These stars were selected to be nonvariable, isolated from their stellar neighbors, located near the center of the field, and bluer than $(B-V)=1.0$. The averaged magnitude error (the standard deviation) for comparison stars was found to be 7.3 mmag in the $V$-band filter. 
 
The candidates for new variable stars were selected from the $V$-band database making use of the analysis of variance method (ANOVA, Schwarzenberg-Czerny 1996). Records in the $B$-band database were used for verifying variability. In total $\sim$8100 light curves covering the magnitude range $9<V<19$ were analyzed.

\section{Multicolor photometry}

\subsection{Cluster structure and CMD morphology} 

\begin{figure}
  \includegraphics[width=7.6cm]{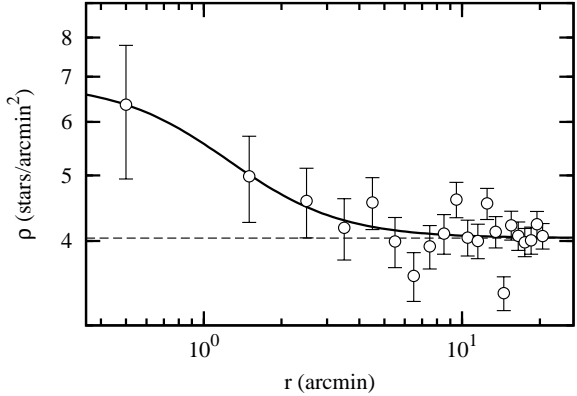}
  \caption{The radial density profile for Trumpler~3. The density uncertainty for each point was estimated assuming the Poisson statistics.}
 \label{fig2}
\end{figure}

\begin{figure*}
 \includegraphics[width=17cm]{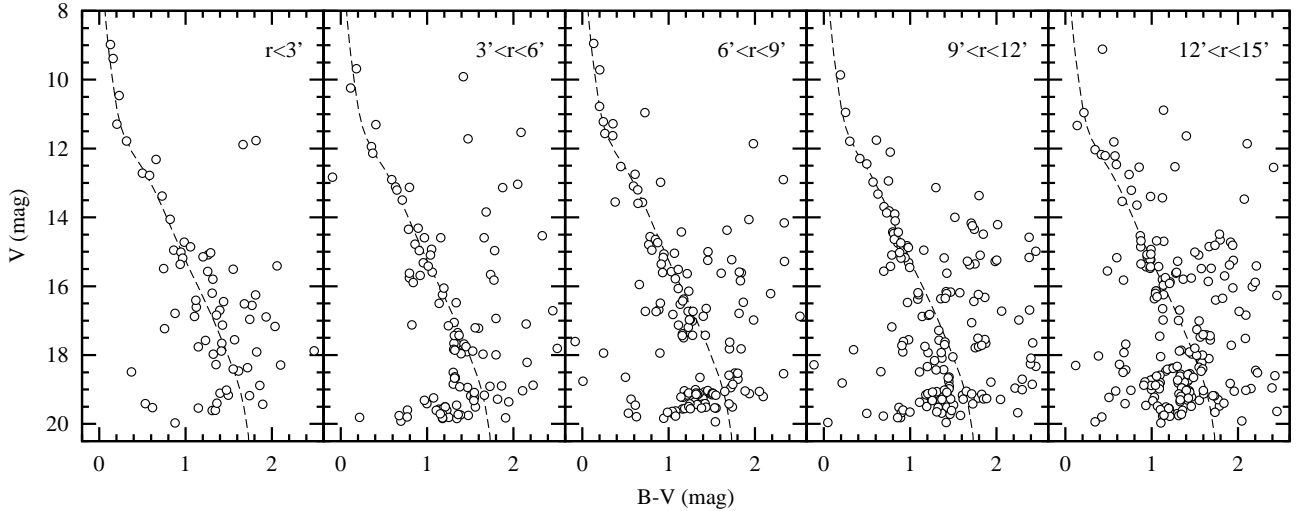}
 \caption{$V$ $vs.$ $(B-V)$ color-magnitude diagrams constructed for successive concentric rings centered on the redetermined cluster coordinates. The decontamination procedure was applied for each diagram. The cluster main sequence can be spotted up to 12'.}
 \label{fig3}
\end{figure*} 

The cluster center coordinates were redetermined using combined photometry. The algorithm was adopted from Maciejewski \& Niedzielski (2007) and started from a tentative position determined by eye. Two perpendicular stripes were cut along declination and right ascension starting from the approximate cluster center and the histogram of star counts was built along each stripe. The bin with the maximum value in both coordinates was taken as a new cluster center. The procedure was repeated iteratively until a stable solution was found. Stars in the cluster field reveal no significant clustering, however the procedure found a stable solution around $\mathrm{RA}=03^{\rm{h}}12.0^{\rm m}$ and $\mathrm{DEC}=+63^{\circ}10'$ ($l=138.\!\!^{\circ}04$, $b=+4.\!\!^{\circ}50$) for the epoch J2000.0. Then, the stellar density inside concentric rings of width $1'$, centered at the redetermined cluster center, was calculated and plotted in a form of the radial density profile (RDP) in Fig.~\ref{fig2}. The data point distribution reveals a stellar overdensity which radius is at least 5'.

To parametrize the density distribution, the King-like function (King 1966) was fitted in a simplified form
\begin{equation}
 \rho(r)=f_{\mathrm{bg}}+\frac{f_{0}}{1+\left(\frac{r}{r_{\mathrm{core}}}\right)^{2}} \,  \;
\end{equation}
(Ka{\l}u\.zny \& Udalski 1992), where $r_{\mathrm{c}}$, $f_{0}$, and $f_{\mathrm{bg}}$ are the core radius, the central density, and the background density level, respectively. We obtained $r_{\mathrm{c}}=1.1\pm0.3$ arcmin, $f_{0}=2.8\pm0.5$ stars/arcmin$^{2}$, and $f_{\mathrm{bg}}=4.04\pm0.08$ stars/arcmin$^{2}$. The fitted profile is sketched with a solid line in Fig.~\ref{fig2} while a dashed line marks a level of $f_{\mathrm{bg}}$.  

Since Trumpler 3 was found to be a low-contrast cluster, a study of the morphology of a color-magnitude diagram (CMD) as a function of an angular distance from the cluster center was carried out to estimate the limiting radius of the cluster $r_{\rm{lim}}$. Fig.~\ref{fig3} presents five CMDs built for concentric, 3'-wide rings, centered on the redetermined cluster center. The background-star contamination was statistically removed with an algorithm adopted from Maciejewski \& Niedzielski (2007). CMDs were built for an investigated region and for an offset field. A concentric offset field of width of $5'$ and starting at $r=16'$ from the cluster center was used. Then CMDs were divided into two-dimensional bins and a number of stars within each box was counted. The cleaned (decontaminated) CMD was built by subtracting a number of stars in an offset box from a number of stars in a corresponding cluster box. The latter number was weighted by the cluster to offset field area ratio. Knowing the number of cluster stars occupying each box, the algorithm randomly chose the required number of stars located in the cluster area and with the adequate magnitude and color index. The width of the concentric rings and the inner radius of the offset field were set after a series of iterative runs. We adopted a value of 3' for the latter quantity as a compromise between the spatial resolution and star numbers. In next step the zero-age main sequence was plotted in each diagram with a dashed line (Fig.~\ref{fig3}). The main sequence was extracted from the Padova isochrones for solar metallicity $Z=0.019$ (Girardi at al. 2002). Then, it was shifted taking into account an interstellar reddening and a distance which were obtained in further steps (see paragraphs 3.2 and 3.3). 

As one can see in Fig.~\ref{fig3}, the cluster main sequence is clearly visible up to fourth area covering $r$ in the range between 9' and 12'. In the last ring ($12'<r<15'$) data point scatter dominates and the presence of the cluster main sequence is disputable. Therefore the value of 12' was adopted as a lower approximation of $r_{\rm{lim}}$.

\subsection{Interstellar reddening}

\begin{figure}
 \includegraphics[width=8.3cm]{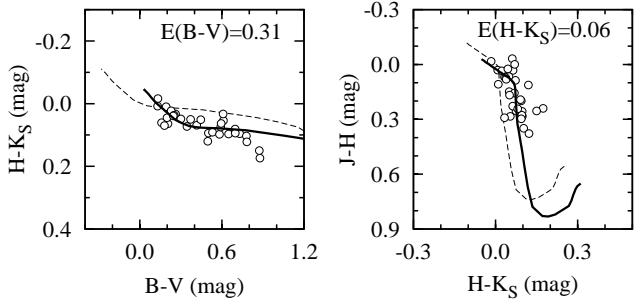}
 \caption{$(H-K_{\mathrm{S}})$ $vs.$ $(B-V)$ and $(J-H)$ $vs.$ $(H-K_{\mathrm{S}})$ diagrams constructed for the Trumpler~3 area. Only stars brighter than $V=14$ mag and located near the main sequence were taken for fitting. The main sequence, fitted by shifting along the reddening vector, is sketched with a continuous line in each diagram while the position of the dereddened main sequence is drawn with a dashed one.}
 \label{fig4}
\end{figure}

$(H-K_{\mathrm{S}})$ $vs.$ $(B-V)$ and $(J-H)$ $vs.$ $(H-K_{\mathrm{S}})$ two color diagrams (TCDs) were constructed to estimate the interstellar reddening toward the cluster. We selected only stars brighter than $V=14$ mag which form the cluster main sequence. The diagrams are plotted in Fig.~\ref{fig4}. The theoretical main sequence was extracted from the Padova isochrones for solar metallicity. The reddenings $E(B-V)=0.31$ and $E(H-K_{\mathrm{S}})=0.06$ were obtained by shifting the theoretical main sequence (a dashed line in Fig.~\ref{fig4}) along the reddening vectors which normal slopes were calculated assuming the universal interstellar extinction law of Schlegel et al. (1998). The location in which a value of $\chi^2$ reached its minimum was taken as a final solution (a continuous line in Fig.~\ref{fig4}). To obtain an independent determination of the color excess in $(B-V)$, the value of $E(H-K_{\mathrm{S}})$ was transformed into $E(B-V)$ applying the relation $\frac{E(H-K_{\mathrm{S}})}{E(B-V)} = 0.209$ taken from Schlegel et al. (1998). We derived $E'(B-V)=0.29$. Both results were found to be consistent with the mean value $\left\langle {E}(B-V) \right\rangle =0.30\pm0.02$.

\subsection{Age and distance}

\begin{figure}
 \includegraphics[width=8.3cm]{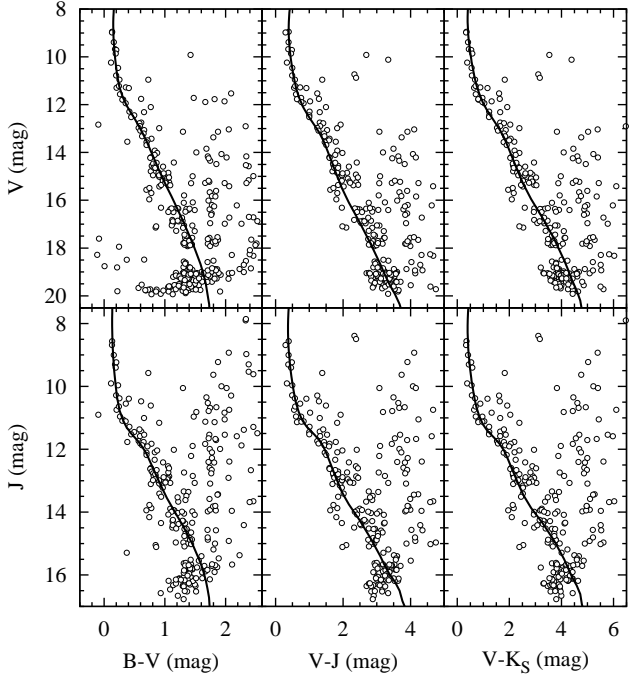}
 \caption{Color-magnitude diagrams constructed for the overall cluster area. The decontamination procedure was applied. The best-fit isochrones are drawn with solid lines.}
 \label{fig5}
\end{figure}

Six CMDs combining optical and near-IR photometry were constructed to determine cluster's age and distance. The consistency of the results was also treated as a verification of the interstellar reddening obtained from the TCDs analysis. $V$ and $J$ magnitudes were combined with $(B-V)$, $(V-J)$, and $(V-K_{\mathrm{S}})$ color indices. Then, the procedure for background-star decontamination (see Sect. 3.1) was run for cluster area within $r_{\rm{lim}}$. The diagrams are plotted in Fig.~\ref{fig5}. 

The age and the distance modulus were derived fitting a set of Padova solar-metallicity isochrones. The fitting algorithm based on the least-squares method. It used stellar magnitudes as weights and was run independently for each diagram. The stars with outstanding colors and magnitudes were rejected manually to increase the reliability of results. The reddenings for individual color indices were fixed and their values were adopted from the analysis of TCDs (see paragraph 3.2). $\left\langle {E}(B-V) \right\rangle$ was transformed into $E(V-J)$ and $E(J-K_{\rm{S}})$ assuming the universal interstellar extinction law of Schlegel et al. (1998). The dereddened distance modulus $(m-M)_{0}$ was calculated under the assumption of the total-to-selective absorption ratio $R=3.315$ (Schlegel et al. 1998). The results of individual fits are collected in Table~1. 

\begin{table}
\centering
\caption{Results of a solar-metallicity isochrone fit for individual CMDs. $E$ denotes the adopted value of the color excess for a given color, and $(m-M)$~-- the fitted value of the apparent distance modulus.} 
\label{tabela1}
\begin{tabular}{l c c c}
\hline
Diagram           & $\log(age)$ & $E$    & $(m-M)$ \\
                  &             &(mag)   & (mag)   \\
\hline 
$V~vs.~(B-V)$          & $7.90$ & $0.30$ & $10.24$ \\
$V~vs.~(V-J)$          & $7.90$ & $0.72$ & $10.23$ \\
$V~vs.~(V-K_{\rm{S}})$ & $7.80$ & $0.88$ & $10.08$ \\
$J~vs.~(B-V)$          & $7.75$ & $0.30$ &  $9.56$ \\
$J~vs.~(V-J)$          & $7.85$ & $0.72$ &  $9.34$ \\
$J~vs.~(V-K_{\rm{S}})$ & $7.80$ & $0.88$ &  $9.45$ \\
\hline
\end{tabular}
\end{table}
 
The final results were calculated as mean values of individual fits and standard deviations were taken as formal errors. We obtained $\log(age)=7.83\pm0.06$, $(m-M)_0=9.2\pm0.1$, and the distance of $0.69\pm0.03$ kpc. The linear diameter was found to be at least $4.8\pm0.2$ pc. While our results seem to be accurate, they may suffer from the systematic errors caused by calibration of photometry and/or by deviation of cluster's metallicity from the solar one.

\subsection{Total mass and number of members}

\begin{figure}
 \includegraphics[width=8.3cm]{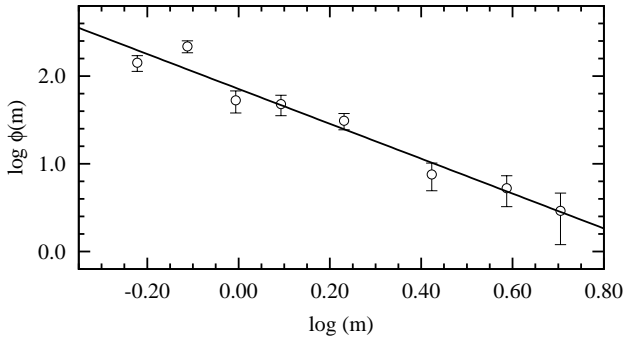}
 \caption{The mass function derived for Trumpler~3. Error bars were calculated assuming the Poisson statistics. The continuous line marks the analytical approximation of the MF.}
 \label{fig6}
\end{figure}

The analysis of a luminosity function (LF) is a tool commonly used for estimating the total mass of a system and the number of its members. To determine these parameters, we used only near-IR photometry and followed the way described in Maciejewski (2008). The LF for overall ($r<r_{\rm{lim}}$) cluster region was built with 1-mag wide bins due to small number of cluster members. The bright end of the LF was determined by the brightest stars while the faint end was set for $J=15.8$ mag -- a value of the 99.9\% Point Source Catalogue Completeness Limit\footnote{following the Level 1 Requirement, according to \textit{Explanatory Supplement to the 2MASS All Sky Data Release and Extended Mission Products} (http://www.ipac.caltech.edu/2mass/releases/allsky/doc)}. The stellar-background LF was built for the offset field ($16'<r<21'$) and then subtracted from the cluster-region LF taking the area proportion into account. The resulting LF was converted into the mass function (MF) using the respective isochrone. The MF is plotted in Fig.~\ref{fig6}. 

The mass function $\phi(m)$ was approximated by a standard relation of the form $\log\phi(m)=-(1+\chi)\log{m}+b_{0}$ where $m$, $\chi$, and $b_{0}$ are the stellar mass, the mass-function slope, and a constant, respectively. As a result of a least-squares fit, we got $\chi=1.0\pm0.2$ which is comparable within error bars to the universal initial mass function (IMF) $\chi_{\rm{IMF}}=1.3 \pm 0.3$ (Kroupa 2001). The fitted relation is drawn with a continuous line in Fig.~\ref{fig6}. 

As a result of extrapolating the MF from the most massive cluster stars down to the H-burning mass limit, the total mass $M_{\rm{tot}}=270\pm40$ $\rm{M}_{\odot}$ and the total number of stars $N_{\rm{tot}}=570\pm90$ were derived. Then, the relaxation time $t_{\rm{relax}}$ and the dynamical-evolution parameter $\tau=\frac{age}{t_{\rm{relax}}}$ were calculated (see Maciejewski \& Niedzielski 2007 and references therein for details of the procedure). We got $\log\tau=0.6$. 

\section{Spectroscopic parallax}

Low-dispersion spectroscopy was used to determine the spectral types of possible bright ($V<12$ mag) cluster's members. Using $V~vs.~(B-V)$ diagram we preselected only stars which form the cluster main sequence. We found 19 such objects. In case of 4 preselected stars their spectra were found to be partially blended with other nearby stars what could make further analysis unreliable. Thus those objects were rejected what limited our sample to 15 stars. Their identifications are given in Table~2.

\begin{table*}
\centering
\caption{Results of spectral studies of individual stars. $Ns$, $d_{\rm{par}}$, and $A_{\rm{V}}$ denote the number of available spectra, the calculated spectroscopic parallax, and the total absorption in the $V$-band filter, respectively. The literature spectral types are given in the column $\rm{Sp_{lit}}$ if available. Types determined by us are listed in the column Sp.} 
\label{tabela2}
\begin{tabular}{l l l c c c c c c c}
\hline
Label & Identification & Coordinates J2000.0 & $\rm{Sp_{lit}}$  & $Ns$ & $V$  & $B-V$ & $d_{\rm{par}}$ & $A_{\rm{V}}$ & Sp \\
      &      &      &      &      & (mag)& (mag) & (kpc)   & (mag) &    \\
\hline 
S01 & GSC 04053-00634 & 031155.8+630840.5 & -- & 1 & 11.78 & 0.31 & 0.73 & 0.78 & A3 \\
S02 & BD+62 528 & 031147.7+631020.1 & B5 &  2 &  8.98 & 0.13 & 1.12 & 1.23 & B2 \\
S03 & BD+62 530 & 031149.7+630812.6 & A0 &  2 &  9.39 & 0.16 & 0.68 & 0.98 & B6.5 \\
S04 & TYC 4053-546-1 & 031141.0+630940.8 & -- &  2 & 11.29 & 0.21 & 0.69 & 0.42 & A3 \\
S05 & GSC 0405300467 & 031205.5+631206.1 & -- &  1 & 10.46 & 0.23 & 1.07 & 1.22 & B6 \\
S06 & BD+62 529 & 031149.6+631341.1 & B5 &  1 &  9.68 & 0.18 & 0.95 & 1.19 & B4 \\
S07 & GSC 04053-00328 & 031207.0+630357.1 & -- &  2 & 11.56 & 0.26 & 0.71 & 0.60 & A3 \\
S08 & HD 19635 & 031256.8+631112.4 & B9 &  1 &  8.95 & 0.13 & 0.73 & 1.03 & B4 \\
S09 & GSC 04053-00576 & 031158.7+630230.7 & -- &  1 & 11.22 & 0.24 & 0.62 & 0.54 & A3 \\
S10 & BD+62 532 & 031238.8+630314.8 & -- &  1 &  9.72 & 0.20 & 0.84 & 1.20 & B5 \\
S11 & TYC 4053-474-1 & 031235.5+631822.2 & -- &  3 & 10.95 & 0.25 & 0.54 & 0.30 & A5.5 \\
S12 & BD+62 534 & 031319.3+631738.9 & A0 &  2 &  9.86 & 0.19 & 0.40 & 0.47 & A2 \\
S13 & BD+62 521 & 030940.3+625914.8 & B5 &  2 &  9.50 & 0.27 & 0.49 & 1.26 & B8 \\
S14 & HD 19354 & 031001.1+632927.5 & A0 &  2 &  9.30 & 0.13 & 0.35 & 0.26 & A2 \\
S15 & GSC 04053-01630 & 030922.7+625300.5 & -- &  1 & 11.50 & 0.24 & 0.71 & 0.52 & A3 \\
\hline
\end{tabular}
\end{table*}


\begin{figure}
 \includegraphics[width=8.3cm]{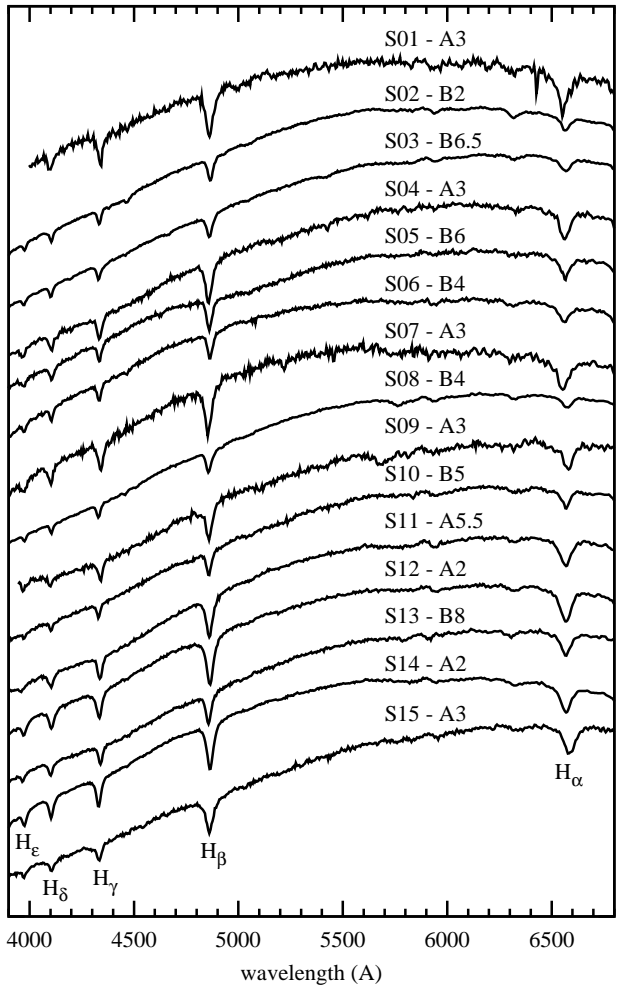}
 \caption{Final spectra of investigated stars. The determined spectral type for each star is given with an identifying label.}
 \label{fig8}
\end{figure}

Two or three spectra were acquired for the majority of stars because individual CCD frames were overlapped. Each spectrum was calibrated to the wavelength using H$_{\alpha}$ -- H$_{\epsilon}$ lines and non-linear calibration equations. Then the spectra were averaged. No flux calibration nor correction for telluric absorption was applied. The final spectra are displayed in Fig.~\ref{fig8} in which they have been shifted vertically for clarity. A few spectra were found to be affected by relatively minor blending which only slightly deformed spectral distribution. This effect was neglected in further analysis.

Spectral classification was done applying criteria from Jaschek \& Jaschek (1987). The sample was found to consist of B and A stars. Thus we used ratios of He 4026, He 4471, K 3933, and Fe 4045 to H$_{\gamma}$ lines. In case of the lower quality spectra we used the equivalent widths of H$_{\beta}$ and H$_{\gamma}$ lines as a main criterion and the presence or absence of weaker characteristic features as an indicator of the spectral type. For two very low quality spectra (S01 and S07) our classification based only on the equivalent widths of hydrogen lines. For brightest stars a set of short- and long-exposure spectra were used to avoid saturation effect for longer wavelengths and to increase the signal-to-noise ratio in the short-wavelength domain. 

\begin{figure}
 \includegraphics[width=8cm]{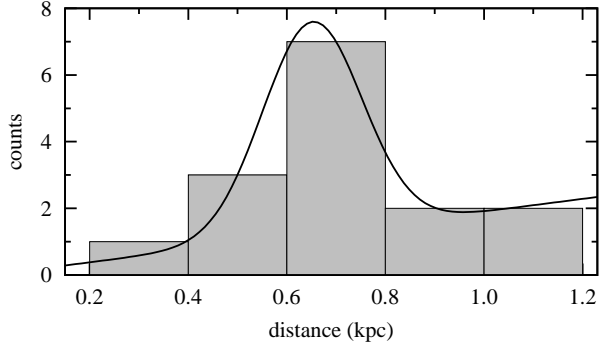}
 \caption{The distribution of spectroscopic parallaxes in the cluster field. The maximum of the fitted Gaussian function is near to 0.7 kpc.}
 \label{fig9}
\end{figure}

From the derived spectral type and measured $(B-V)$ we calculated $E(B-V)$, making use of tabulated calibrations of the intrinsic color indices $(B-V)_0$ available for each spectral type. Assuming the total-to-selective absorption ratio of $R=3.315$ (Schlegel et al. 1998), the total absorption in the $V$-band filter $A_{\rm{V}}$ was calculated. The $V$ magnitude, corrected for reddening, was used to derive the distance modulus and then the spectroscopic parallaxes $d_{\rm{par}}$. The results for individual stars and literature spectral types, if available, are collected in Table~2. The distribution of derived distances is presented in Fig.~\ref{fig9} in a form of a column graph. The maximum of the distribution is clearly visible around 0.7 kpc. To characterize the distribution, a Gaussian function with a background gradient was fitted with the least-squares method. As a result, the distance to the cluster was found to be $0.65\pm0.09$ kpc where $\sigma$ of the Gaussian distribution was taken as the error. 

\begin{figure}
 \includegraphics[width=7.5cm]{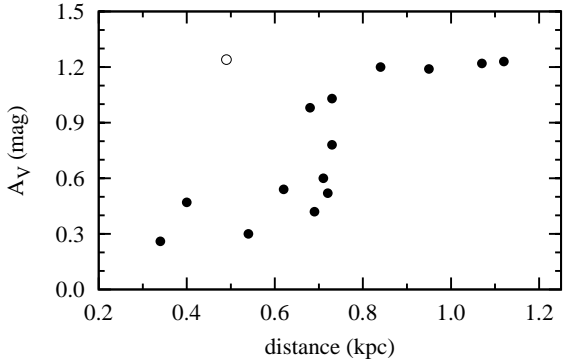}
 \caption{The distribution of the total absorption in the $V$ band $A_{\rm{V}}$ for stars investigated spectroscopically as a function of their spectroscopic parallax. The outstanding point, marked with an open circle, is a foreground star BD+62~521.}
 \label{fig10}
\end{figure}

The relation of $A_{\rm{V}}$ $vs.$ $d_{\rm{par}}$ is plotted in Fig.~\ref{fig10} for individual stars. For $d_{\rm{par}}<0.7$ kpc the absorption increases linearly. Around $d_{\rm{par}}=0.7$ (the distance of the cluster) a rapid growth or discontinuity is clearly visible. For $d_{\rm{par}}>0.7$ kpc the absorption increases linearly again. This relation of a step-like shape suggests that a bubble of interstellar matter is associated with the cluster.

\section{Variable stars}

The goal of our wide-field monitoring was to detect short period variable stars -- members of Trumpler~3. Their variability could be caused by stellar pulsations or eclipses of binary star components. Further analysis of their properties was expected to enrich studies of the cluster. Since the cluster was found to be relatively young, no long-period regular or semi-regular variable members were expected. 

As a result of our survey, 24 variable stars were detected. The preliminary variability type of each one was based on its light-curve morphology, period $P$, and amplitude of variance. We distinguished (i) three types of eclipsing binaries: contact systems of W~UMa type (EW), semi-detached systems of $\beta$~Lyr type (EB), detached systems of Algol type (EA), (ii) regular pulsating variables: $\delta$ Sct variables (DSCT), RR Lyr stars (RR), Cepheids (DCEP), and (iii) miscellaneous variables (MISC) including semi-regular pulsators (SR).

In case of non-eclipsing variables their $(J-H)$ and $(H-K_{\rm{S}})$ color indices were calculated and stars were placed in the $(J-H)$ $vs.$ $(H-K_{\rm{S}})$ and $\log P$ $vs.$ $(J-H)$ diagrams (Pojma\'nski \& Maciejewski 2004). This step allowed us to verify the classification and in some cases helped us to remove the RR/EW and DCEP/SR degeneracy.

The light curves of about two dozens of brightest stars which turned out to be saturated in short exposures were extracted from the Northern Sky Variability Survey (NSVS, Wo\'zniak et al. 2004) and then analyzed for variance. Only one from them -- TYC 4053-630-1 -- was found to reveal low-amplitude, periodic changes in brightness. As the star was found to be an unlikely cluster member and its spectrum shows structures typical for chemically peculiar stars, we do not present its studies in this paper. The dedicated spectroscopic and photometric investigation of this object will be presented elsewhere.

\subsection{Eclipsing binaries}

\begin{figure}
 \includegraphics[width=8.3cm]{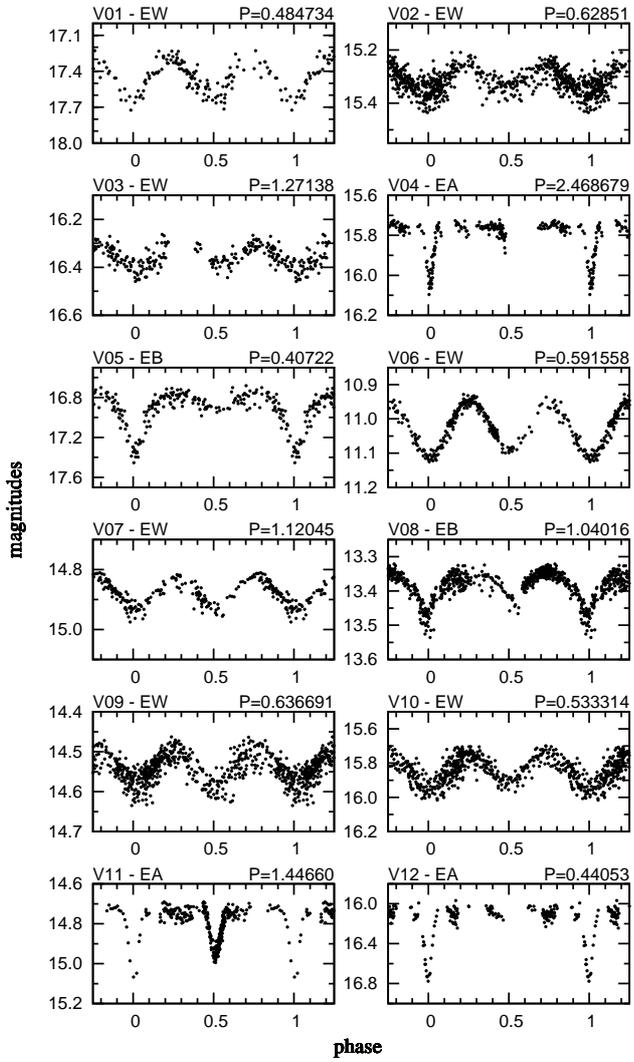}
 \caption{$V$-band light curves of eclipsing binaries discovered in the field of Trumpler~3.}
 \label{fig11}
\end{figure}

Twelve eclipsing binaries were discovered in the field of Trumpler~3. They are listed in Table~3 in which parameters of their light curves are given. Phased light curves in the $V$-band filter are plotted in Fig.~\ref{fig11}. Three of them are located within the cluster limiting radius. The remaining ones may be treated as stars of the Galactic background. We found 7 contact binaries (EW), 3 detached systems (EA), and two semi-detached ones (EB). The individual types constitute 58\%, 25\%, and 17\% of the sample, respectively, what is close to a typical distribution of the Galactic field (Pojma\'nski 2002).    

\begin{table*}
\centering
\caption{The list of eclipsing binaries. The columns contain: object's label, an identification, coordinates, the distance from the cluster center  $r$ in arcmin, the maximal brightness in the $V$-band filter $V_{\rm{max}}$ in magnitudes, the color index in maximum  $B-V$, the depth of a primary minimum in $V$ and $B$ bands $\Delta V_{\rm{pri}}$ and $\Delta B_{\rm{pri}}$, respectively, the depth of a secondary minimum in $V$ band $\Delta V_{\rm{sec}}$, the period of variance $P$,  the epoch of a primary minimum $T_{0}$, and the type of variability.} 
\label{tabela3}
\begin{tabular}{l l c c c c c l}
\hline
Label & Identification & $r$ & $V_{\rm{max}}$ & $\Delta V_{\rm{pri}}$ & $\Delta V_{\rm{sec}}$ & $P$ & Type\\
      & Coordinates J2000.0 &  & $B-V$ & $\Delta B_{\rm{pri}}$ & & $T_0$ &    \\
\hline 
V01 & GSC2.3 NAVR015760 & 7.7  & 17.27 & 0.39 & 0.37 & $0.484734\pm0.000012$ & EW \\
    & 031307.0+631128.9 &      &  1.15 & 0.49 &      & 2453651.0429          &    \\
V02 & GSC2.3 NAVR015760 & 9.3  & 15.25 & 0.11 & 0.08 & $0.62851\pm0.00006$   & EW \\
    & 031049.1+630518.1 &      &  0.96 & 0.13 &      & 2453652.0799          &    \\
V03 & GSC2.3 NAVO004630 & 10.6 & 16.28 & 0.13 & 0.11 & $1.27138\pm0.00010$   & EW \\
    & 031044.1+631619.2 &      &  1.10 & 0.13 &      & 2453653.5731          &    \\
V04 & GSC2.3 NAVR013234 & 13.4 & 15.72 & 0.34 &  ?   & $2.468679\pm0.000082$ & EA \\
    & 031335.4+630208.8 &      &  0.90 & 0.30 &      & 2453654.2974          &    \\
V05 & GSC2.3 NAVO004282 & 16.6 & 16.74 & 0.65 & 0.16 & $0.407224\pm0.000010$ & EB \\
    & 030939.3+631453.8 &      &  1.25 & 0.63 &      & 2453650.8781          &    \\
V06 & GSC 04053-00029   & 20.8 & 10.94 & 0.17 & 0.16 & $0.591558\pm0.000018$ & EW \\
    & 031337.7+632741.1 &      &  1.04 & 0.19 &      & 2453672.0377          &    \\
V07 & GSC2.3 NAVO012804 & 21.4 & 14.82 & 0.12 & 0.11 & $1.12045\pm0.00015$   & EW \\
    & 031214.9+633120.0 &      &  1.00 & 0.13 &      & 2453653.5633          &    \\
V08 & GSC 04053-00275   & 22.5 & 13.34 & 0.13 & 0.09 & $1.04016\pm0.00013$   & EB \\
    & 031306.3+624849.9 &      &  0.81 & 0.15 &      & 2453651.3031          &    \\
V09 & GSC 04053-00361   & 23.5 & 14.49 & 0.10 & 0.10 & $0.636691\pm0.000055$ & EW \\
    & 031425.8+625319.6 &      &  1.01 & 0.08 &      & 2453651.9990          &    \\
V10 & GSC2.3 NAVP031574 & 29.1 & 15.74 & 0.20 & 0.18 & $0.533314\pm0.000036$ & EW \\
    & 031348.0+624341.6 &      &  1.50 & 0.19 &      & 2453651.3989          &    \\
V11 & GSC 04053-00046   & 33.8 & 14.73 & 0.34 & 0.24 & $1.44660\pm0.00003$   & EA \\
    & 030934.7+633941.8 &      &  1.14 & 0.30 &      & 2453651.8018          &    \\
V12 & GSC2.3 NAVO020344 & 35.1 & 16.06 & 0.72 &  ?   & $2.50240\pm0.00025$   & EA \\
    & 031350.4+634255.1 &      &  1.05 & 0.71 &      & 2453652.3477          &    \\
\hline
\end{tabular}
\end{table*}


\subsection{Regular pulsating variables}

\begin{figure}
 \includegraphics[width=8.3cm]{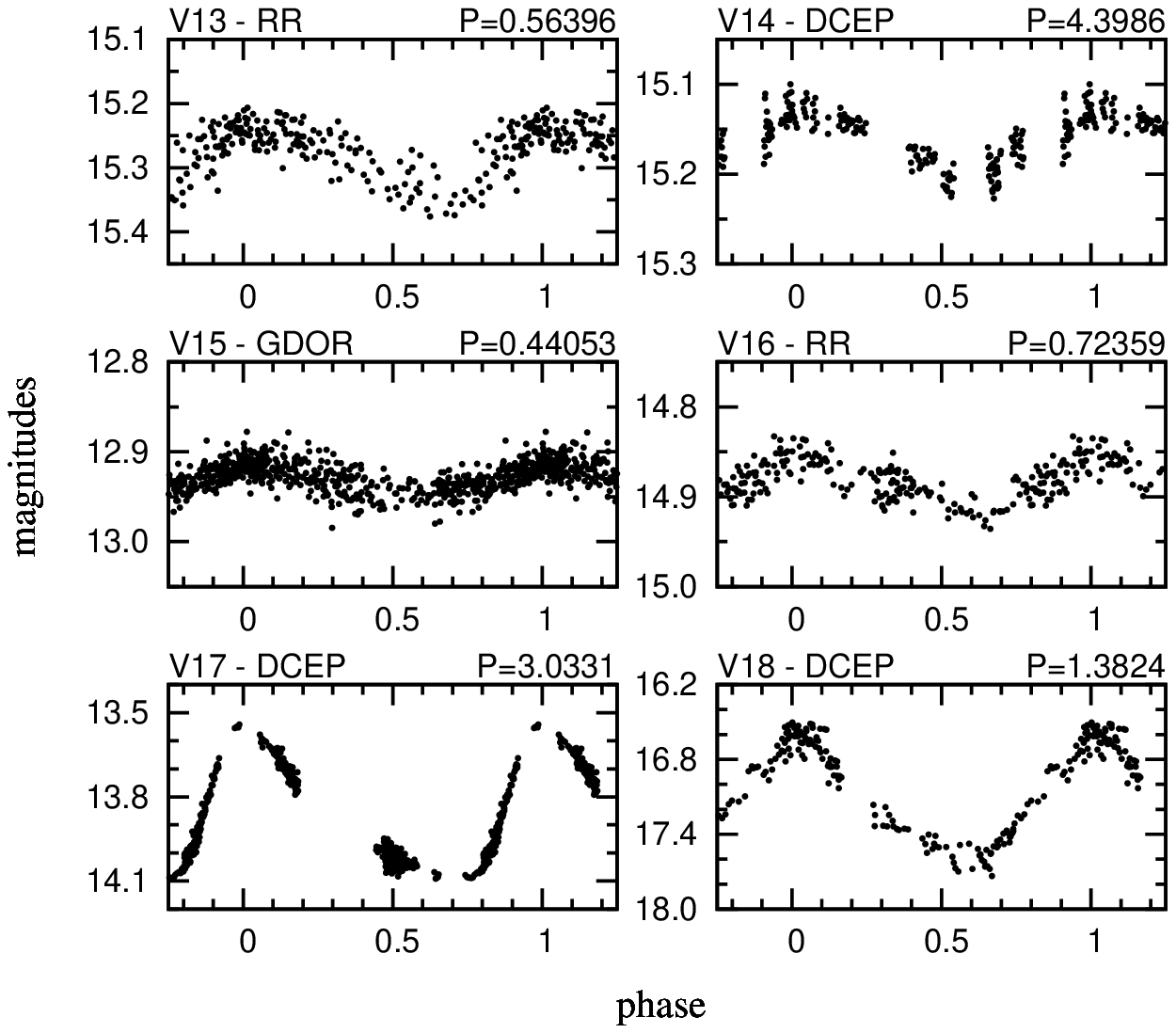}
 \caption{$V$-band light curves of regular pulsating variables in the field of Trumpler~3.}
 \label{fig12}
\end{figure}

We found 6 regular pulsating variables: 2 of the RR type, a $\gamma$ Dor star (GDOR), and 3 Cepheids. Four stars are potential cluster members due to their location on the sky. Table~4 contains parameters of new pulsating stars and their phased light curves are displayed in Fig.~\ref{fig12}. For each star the interval between phases of its maximum and minimum $\Delta \phi$ was determined to characterize the asymmetry of a light curve.

\begin{table*}
\centering
\caption{The list of regular pulsating variables. The columns contain: object's label, an identification, coordinates, the distance from the cluster center  $r$ in arcmin, the maximal brightness in the $V$-band filter $V_{\rm{max}}$ in magnitudes, the color index in maximum  $B-V$, the amplitude of variance in $V$ and $B$ bands $\Delta V$ and $\Delta B$, respectively, the interval between phases of a maximum and a minimum $\Delta \phi$, the period of variance $P$,  the epoch of a maximum $T_{0}$, and the type of variability.} 
\label{tabela4}
\begin{tabular}{l l c c c c c l}
\hline
Label & Identification & $r$ & $V_{\rm{max}}$ & $\Delta V$ & $\Delta \phi$ & $P$ & Type\\
      & Coordinates J2000.0 &  & $B-V$ & $\Delta B$ & & $T_0$ &    \\
\hline 
V13 & GSC2.3 NAVR022373 &  6.6 & 15.23 & 0.14 & 0.69 & $0.56396\pm0.00002$   & RR \\
    & 031230.2+631541.0 &      &  0.93 & 0.15 &      & 2453652.1617          &    \\
V14 & GSC2.3 NAVR011576 & 10.1 & 15.12 & 0.09 & 0.60 & $4.399\pm0.001$   & DCEP \\
    & 031153.3+625956.2 &      &  0.88 & 0.10 &      & 2453663.1388          &    \\
V15 & GSC 04053-00233   & 10.2 & 12.91 & 0.05 & 0.58 & $0.44053\pm0.00002$   & GDOR \\
    & 031030.0+630911.4 &      &  0.63 & 0.05 &      & 2453663.1388          &    \\
V16 & GSC2.3 NAVO005536 & 11.1 & 14.85 & 0.07 & 0.65 & $0.72359\pm0.00053$   & RR \\
    & 031110.2+631934.3 &      &  0.87 & 0.07 &      & 2453651.6810          &    \\
V17 & GSC 04053-00115   & 16.3 & 13.56 & 0.51 & 0.76 & $3.0331\pm0.0009$   & DCEP \\
    & 031347.1+632058.0 &      &  1.21 & 0.65 &      & 2453654.8598          &    \\
V18 & GSC2.3 NAVO020013 & 33.6 & 16.60 & 0.99 & 0.64 & $1.3824\pm0.0002$   & DCEP \\
    & 031307.0+634243.5 &      &  1.29 & 1.18 &      & 2453653.0144          &    \\
\hline
\end{tabular}
\end{table*}


\subsection{Miscellaneous variables}
 
\begin{figure}
 \includegraphics[width=8.3cm]{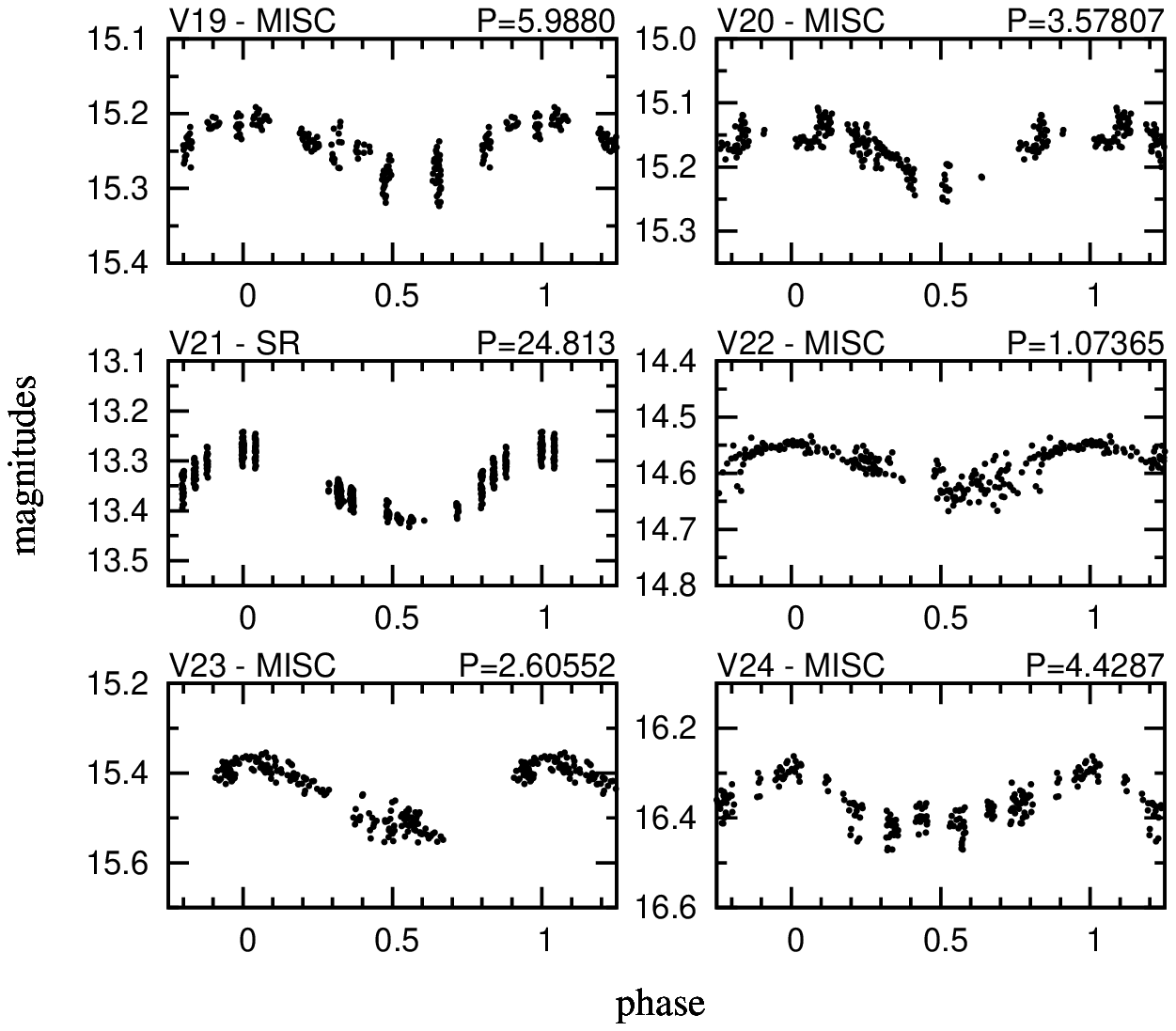}
 \caption{$V$-band light curves of miscellaneous variables discovered in the field of Trumpler~3. }
 \label{fig13}
\end{figure}

Six variable stars were found to have sinusoid-like light curves with periods longer than one day and amplitudes smaller than 0.2 mag in $V$. They are collected in Table~5 and their light curves are plotted in Fig.~\ref{fig13}. Star V21 is probably a SR variable due to its location in the $(J-H)$ $vs.$ $(H-K_{\rm{S}})$ and $\log P$ $vs.$ $(J-H)$ diagrams. The nature of brightness changes of remaining stars is not clear, thus we classified them as miscellaneous variables. 
 
\begin{table*}
\centering
\caption{The list of miscellaneous variables. The columns contain: object's label, an identification, coordinates, the distance from the cluster center $r$ in arcmin, the maximal brightness in the $V$-band filter $V_{\rm{max}}$ in magnitudes, the color index in maximum  $B-V$, the amplitude of variance in $V$ and $B$ bands $\Delta V$ and $\Delta B$, respectively, the period of variance $P$,  the epoch of a maximum $T_{0}$, and the type of variability.} 
\label{tabela5}
\begin{tabular}{l l c c c c l}
\hline
Label & Identification & $r$ & $V_{\rm{max}}$ & $\Delta V$ & $P$ & Type\\
      & Coordinates J2000.0 &  & $B-V$ & $\Delta B$ & $T_0$ &    \\
\hline 
V19 & GSC2.3 NAVR016793 &  4.5 & 15.21 & 0.08 & $5.9880\pm0.0051$     & MISC \\
    & 031132.9+630643.8 &      &  1.03 & 0.11 & 2453666.1386          &    \\
V20 & GSC2.3 NAVR021660 & 11.2 & 15.13 & 0.09 & $3.57807\pm0.00075$   & MISC \\
    & 031331.1+631429.8 &      &  0.90 & 0.09 & 2453661.5354          &    \\
V21 & GSC 04053-00388   & 16.9 & 13.28 & 0.14 & $24.83\pm0.03$        & MISC/SR \\
    & 031114.5+625356.1 &      &  1.93 & 0.18 & 2453709.1530          &    \\
V22 & GSC 04053-00069   & 17.2 & 14.55 & 0.09 & $1.07365\pm0.00017$   & MISC \\
    & 031204.4+632713.1 &      &  1.02 & 0.13 & 2453653.1765          &    \\
V23 & GSC2.3 NAVR016193 & 19.9 & 15.38 & 0.17 & $2.60552\pm0.00074$   & MISC \\
    & 031452.6+630600.2 &      &  1.29 & 0.19 & 2453655.2693          &    \\
V24 & GSC2.3 NAVR004948 & 25.2 & 16.29 & 0.14 & $4.4287\pm0.0026$     & MISC \\
    & 031436.8+625217.3 &      &  1.10 & 0.15 & 2453660.2141          &    \\
\hline
\end{tabular}
\end{table*}

\subsection{Membership}

\begin{figure}
 \includegraphics[width=6cm]{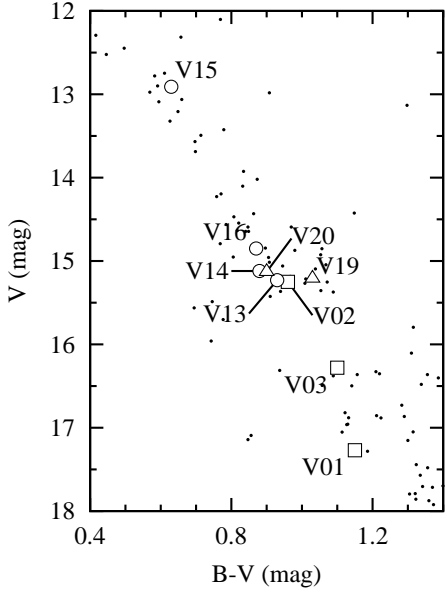}
 \caption{The color-magnitude diagram for Trumpler~3 with marked
individual variables situated within the cluster limiting radius.}
 \label{fig14}
\end{figure}

To investigate the membership likelihood of potential cluster variables, they were plotted in the cluster CMD, as it is shown in Fig.~\ref{fig14}. As one can note, all of them seem to be located near the cluster main sequence. Six variables of various types are clustered around $V=15$ mag.

Variables V01, V02, and V03 are short period contact binaries. To verify their membership, their absolute magnitudes may be derived in two ways under the assumption that these variables belong to the cluster. The absolute magnitude $M_{\mathrm{V}}^{\mathrm{iso}}$ of the system may be calculated from its maximum brightness $V_{\mathrm{max}}$ and cluster's distance modulus $(m-M)_{\rm{V}}$ as 
\begin{equation}
M_{\mathrm{V}}^{\mathrm{iso}}=V_{\mathrm{max}} - (m-M)_{\rm{V}}\, . \;
\end{equation}
On the other hand, the absolute magnitude $M_{\mathrm{V}}^{\mathrm{EW}}$ can also be obtained from the empirical formula 
\begin{equation}
    M_{\mathrm{V}}^{\mathrm{EW}}=-4.44 \log P + 3.02(B-V)_{0} + 0.12 \, , \;
\end{equation}
where $P$ is the period of variation in days and $(B-V)_{0}$ is the dereddened color index (Ruci\'nski \& Duerbeck 1997, Ruci\'nski 2004). If a system belonged to the cluster both values would be identical within a typical error of 0.25 mag (Ruci\'nski 2004). The first method gives $M_{\mathrm{V}}^{\mathrm{iso}}$ of 7.1, 5.1, and 6.1 mag for V01, V02, and V03, respectively, while the second one returns $M_{\mathrm{V}}^{\mathrm{EW}}$ of 4.1, 3.0, and 2.1 mag, respectively. In all three cases the discrepancy of results is significant what suggests that none of contact systems belongs to the cluster.

Pulsating variables of RR Lyr type and Cepheids are not expected in such clusters as Trumpler~3, thus V13, V14, and V16 may be treated as objects of the Galactic background. Star V15 was classified as a $\gamma$~Dor variable. Assuming it belongs to the cluster, its absolute magnitude would be $\sim$2.7 mag, what is in the range of typical values of $\gamma$~Dor variables (see e.g. Henry \& Fekel 2002). That allows to treat V15 as a likely cluster member.

Stars V19 and V20 are variables of unknown type. They reveal brightness variation with amplitudes smaller than 0.1 mag and period of a few days. Assuming they belong to the cluster, their absolute magnitudes would be $\sim$5 mag. Since no variability caused by pulsations is expected in this absolute magnitude range, the membership of both stars seems to be unlikely.

\section{Summary}
 
The open cluster Trumpler 3 was found to be a low contrast stellar ensemble with the limiting radius at least 12' which is a value between results of Dias et al. (2002) and Kharchenko at al. (2005). Our value results in the cluster linear radius of 4.8 pc. 

Our photometric studies indicate that the interstellar reddening toward the cluster is greater than Kharchenko at al. (2005) reported and was found to be $E(B-V)=0.30\pm0.02$ mag. The distribution of interstellar extinction in the cluster field suggests the existence of a bubble of interstellar matter associated with the cluster. 

The distance determined from photometric studies was found to be $0.69\pm0.03$ kpc and was found to be consistent with the spectroscopic parallax of $0.65\pm0.09$ kpc. Our results are consistent with the value given by Trumpler (1930). The distance given by Kharchenko at al. (2005) seems to be underestimated. The cluster is located in the outer edge of the Orion Spur. 

Trumpler~3 was found to be younger than studies of Kharchenko at al. (2005) indicated. Its age was found to be $70\pm10$ Myr ($\log(age)=7.83\pm0.06$). The mass function distribution was found to be close to the universal IMF. The total mass was estimated to be $M_{\rm{tot}}=270\pm40$ $\rm{M}_{\odot}$ and the total number of stars $N_{\rm{tot}}=570\pm90$. The mean stellar density within the cluster volume is 1.2 star per pc$^3$. The cluster turned out to be older than its relaxation time but no sight of evaporation of low-mass stars was detected. 

As a result of the wide-field search for variable stars we found 24 variables. Only one of them -- a $\gamma$~Dor variable -- was found to be a likely cluster member.

\textit{Acknowledgements}:
This research has made use of the SIMBAD data base and is partially supported by UMK grant 411-A. GM acknowledges support from the EU in the FP6MC ToK project MTKD-CT-2006-042514. {\L}B acknowledges support from the \textit{Sty-pendia dla doktorantow 2008/2009} -– ZPORR SPS.IV-3040-UE/ 204/2009. This publication makes use of data products from the Two Micron All Sky Survey, which is a joint project of the University of Massachusetts and the Infrared Processing and Analysis Center/California Institute of Technology, funded by the National Aeronautics and Space Administration and the National Science Foundation.


\begin{thebibliography}{}
  \bibitem{} Barnard, E.~E., Frost, E.~B., Calvert, M.~R.: 1927, A Photographic Atlas of Selected Regions of the Milky Way, Carnegie Institution of Washington
  \bibitem{} Dias, W.~S., Alessi, B.~S., Moitinho, A., Lepine, J.~R.~D: 2002, A\&A, 389, 871
  \bibitem{} Girardi, L., Bertelli, G., Bressan, A., et al.: 2002, A\&A, 391, 195
  \bibitem{} Henry, G.~W., Fekel, F.~C.: 2002, PASP, 114, 988
  \bibitem{} Jaschek, C., Jaschek, M.: 1987, The classification of stars, Cambridge University Press
  \bibitem{} Ka{\l}u\.zny, J., Udalski, A.: 1992, AcA, 42, 29
  \bibitem{} Kharchenko, N.~V., Piskunov, A.~E., R\"oser, S., Schilbach, E., Scholz, R.-D.: 2005, A\&A, 438, 1163
  \bibitem{} King, I.: 1966, AJ, 71, 64
  \bibitem{} Kroupa, P.: 2001, MNRAS, 322, 231
  \bibitem{} Landolt, A.~U.: 1992, AJ, 104, 340
  \bibitem{} Maciejewski, G.: 2008, Acta Astron., 58, 389
  \bibitem{} Maciejewski, G., Niedzielski, A.: 2007, A\&A, 467, 1065
  \bibitem{} Mermilliod, J.C.: 1996, in The Origins, Evolution and Destinies of Binary Stars in Cluster, ed. E. F. Milone \& J.-C. Mermilliod, ASP Conf. Ser. 90, 475  
  \bibitem{} Mermilliod, J.~C., Mayor, M., Udry, S.: 2008, A\&A, 485, 303
  \bibitem{} Niedzielski, A., Maciejewski, G., Czart, K.: 2003, Acta Astron., 53, 281
  \bibitem{} Pojma\'nski, G.: 2002, Acta Astron., 52, 397
  \bibitem{} Pojma\'nski, G., Maciejewski, G.: 2004, Acta Astron., 54, 153
  \bibitem{} Ruci\'nski, S.~M.: 2004, New Astronomy Reviews, 48, 703
  \bibitem{} Ruci\'nski, S.~M., Duerbeck,~H.~W.: 1997, PASP, 109, 1340
  \bibitem{} Ruprecht, J.: 1966, BAICz, 17, 33
  \bibitem{} Schlegel, D.~J., Finkbeiner, D.~P., Davis, M.: 1998, ApJ, 500, 525
  \bibitem{} Stetson, P.~B.: 1987, PASP, 99, 191
  \bibitem{} Strutskie, M.~F., Cutri, M.~F., Stiening, R., et al.: 2006, AJ, 131, 1163
  \bibitem{} Schwarzenberg-Czerny, A.: 1996, ApJ, 460, 107
  \bibitem{} Trumpler, R.~J.: 1930, Lick Observ. Bull., 14, 154
  \bibitem{} Wo\'zniak, P.~R., Vestrand, W.~T.; Akerlof, C.~W., et al.: 2004, AJ, 127, 2436
\end{thebibliography}
\end{document}